\documentclass[english]{article}
\usepackage{lmodern}

\usepackage[T1]{fontenc}
\usepackage[latin9]{inputenc}
\usepackage{babel}
\usepackage{verbatim}
\usepackage{varioref}
\usepackage{endnotes}
\usepackage{amsmath}
\usepackage[unicode=true,pdfusetitle,
 bookmarks=true,bookmarksnumbered=false,bookmarksopen=false,
 breaklinks=true,pdfborder={0 0 0},pdfborderstyle={},backref=false,colorlinks=false]
 {hyperref}

\makeatletter
\providecommand*{\code}[1]{\texttt{#1}}

\@ifundefined{date}{}{\date{}}
\makeatother

\begin{document}
\title{Using Category Theory in Modeling Generics\\
in Object-Oriented Programming\thanks{This extended abstract has been accepted for presentation at the Applied
Category Theory (ACT 2019) conference organized by the Department
of Computer Science at Oxford University, London, UK on July 15-19\protect\textsuperscript{th},
2019.}}
\author{Moez A. AbdelGawad\\
Informatics Research Institute, SRTA-City, Alexandria, Egypt\\
\texttt{moez@cs.rice.edu}}
\maketitle
\begin{abstract}
Modeling generics in object-oriented programming languages such as
Java and C\# is a challenge. Recently we proposed a new order-theoretic
approach to modeling generics. Given the strong relation between order
theory and category theory, in this extended abstract we present how
also some tools from category theory, such as adjunctions, monads
and operads, are used in our approach to modeling generics.
\end{abstract}

\paragraph{Introduction}

Support for generic classes and generic interfaces (called \emph{generics})
was added to Java to enhance the expressiveness of its type system~\cite{JLS05,JLS18,Bracha98,Corky98,Thorup99,GenericsFAQWebsite}.
Generics are supported in other mainstream nominally-typed object-oriented
programming (OOP) languages, such as C\#~\cite{CSharp2015}, Scala~\cite{Odersky14},
C++~\cite{CPP2017}, and Kotlin~\cite{Kotlin18}, for the same purpose.
However, in spite of much research, the simple and accurate mathematical
modeling of generics in nominally-typed OOP languages remains a challenge,
mainly due to the roughness of some features of generics such as variance
annotations (\emph{e.g.}, Java wildcards), $F$-bounded generics%
, and (Java) erasure~\cite{Torgersen2004,MadsTorgersen2005,Cameron2007,Cameron2008,Summers2010,Tate2011}.

In a recent paper~\cite{AbdelGawad2019g} %
we outlined a new approach, based on order theory, for modeling generics
in OOP languages. In the order-theoretic approach to modeling generics
we use concepts such as products of posets, intervals over posets,
and pre-/post-fixed points. Details of different parts of this approach
are presented in~\cite{AbdelGawad2018a,AbdelGawad2018b,AbdelGawad2018c,AbdelGawad2018e,AbdelGawad2019b}.

Most concepts in order theory have more general counterparts in category
theory~\cite{Fong2018,Priestley2002,spivak2014category}. These include
Galois connections which, in category theory, are generalized to adjunctions,
and closure and kernel operators which are generalized to monads and
comonads. As such, in the order-theoretic approach to modeling generics
we also make use of some concepts from category theory. Further, some
useful tools in category theory, such as operads, either have trivial
or no clear counterparts in order theory. In this abstract we present
a brief summary of how, in particular, we use adjunctions, monads
(together with algebras and coalgebras), and operads in our approach
to model generics in object-oriented programming.

\paragraph{The Java Erasure Adjunction and Free Types}

All nominally-typed OOP languages, such as Java, C\#, C++, Scala and
Kotlin, are class-based. Class-based OOP languages that are statically-typed
have two fundamental ordering relations: the \emph{subclassing }(also
called \emph{inheritance}) relation between classes, and the \emph{subtyping}
relation between reference types\endnote{\emph{To shorten this extended abstract, some technical details are
left out }(\emph{and online preprints that include these details are
cited})\emph{, while some other details }(\emph{that are not available
online})\emph{ and some of the lesser-significant text is moved to
these endnotes. These endnotes are included here as an appendix mainly
for ACT'19 reviewers to make a more in-depth assessment of the approach
we present in this abstract. As such, these endnotes are not intended
for inclusion in the final version of the abstract, and can be included
only in a full preprint version available online }(e.g.\emph{, on
arXiv})\emph{.}\medskip{}
}\textsuperscript{,}\endnote{In our work interfaces and traits are treated as abstract classes.
The term `classes' here thus refers to classes and similar (reference-)type-constructing
constructs in OOP such as interfaces and traits. Also, reference types
are sometimes called \emph{object types}, \emph{class types}, just
\emph{types}, or (when generics are supported) \emph{generic types
}or \emph{parameterized types}.\medskip{}
}. In the order-theoretic approach to modeling generics, %
the subtyping relation in nominally-typed OOP is constructed iteratively,
solely from the subclassing relation, using novel order-theoretic\emph{
}operators developed particularly for this purpose~\cite{AbdelGawad2018a,AbdelGawad2018b,AbdelGawad2018c}.

Maintaining a clear distinction between the subclassing and subtyping
relations also enables recognizing and defining an adjunction between
the two relations, when the two ordered sets corresponding to the
two relations are viewed as categories rather than posets. We call
this adjunction the \emph{Java Erasure Adjunction} (\emph{JEA}).\endnote{Although \emph{JEA} involves category-theoretic concepts such as adjoints
and free types, but the basis of the adjunction is order-theoretic.
Also, while erasure is discussed frequently in relation to Java (to
support migration-compatibility), but erasure is not specific to Java
and, as a mapping between classes and types, it is applicable to any
generic OOP language. As such, the \emph{JEA} may also be called the
\emph{Erasure Galois Connection} (\emph{EGC}).\medskip{}
} The left adjoint of \emph{JEA} is\emph{ }Java erasure, which ``erases''
type arguments of a parameterized type~\cite{JLS18}, and as such
maps a parameterized type to a class. The right adjoint of the \emph{JEA}
adjunction is the newly-named notion of a \emph{free type} corresponding
to a class, which maps any class to the type expressing the ``most
general wildcard instantiation'' of the class (\emph{e.g.}, in Java,
a generic class \code{C} with one type parameter has the parameterized
type \code{C<?>} as its corresponding free type).

As for any adjunction, to properly define an adjunction the two maps
of \emph{JEA} have to work in tandem to satisfy a preservation condition.
In particular, if $E$ is the erasure functor and $FT$ is the free
type functor, and if $\le$ denotes the subclassing relation and $<:$
denotes the subtyping relation, then for $E$ and $FT$ to define
an adjunction we must have 
\begin{equation}
E(t)\le c\Longleftrightarrow t<:FT(c),\label{eq:adj}
\end{equation}
for all types $t$ and classes $c$.

In words, this condition says that the erasure $E(t)$ of a parameterized
type $t$ is a subclass of class $c$ if and only if $t$ is a subtype
of the free type $FT(c)$ corresponding to class $c$. This is a true
statement in generics,\endnote{Consider, for example, the statement (in Java) 
\[
\mathtt{LinkedList\le List\Longleftrightarrow LinkedList\negthickspace<\negthickspace T\negthickspace>\;<:\;List\negthickspace<?\negthickspace>}
\]
where, in Equation~(\ref{eq:adj})\vpageref{eq:adj}, type variable
$t$ is instantiated to the generic type \code{LinkedList<T>} for
all type arguments $\mathtt{T}$ (\emph{e.g.}, \code{String} or \code{Integer}
or \code{?~extends Number}) and class variable $c$ is instantiated
to class \code{List}. This statement asserts that class \code{LinkedList}
in Java is a subclass of \code{List} if and only if all instantiations
of \code{LinkedList} are subtypes of the free type \code{List<?>}---which
is a true statement in Java.\medskip{}
}\textsuperscript{,}\endnote{\emph{Digressive Note}:\emph{ }The preservation condition expressed
by Equation~(\ref{eq:adj})\vpageref{eq:adj} is equivalent to the
statement stating that, for any two classes \code{C} and \code{D},
if \code{D} is a subclass of (\emph{i.e.}, inherits from) \code{C}
then all parameterized types that are instantiations of \code{D},
and their subtypes, are subtypes of the free type \code{C<?>} corresponding
to class \code{C} \emph{and} vice versa, \emph{i.e.}, if all instantiations
of some class \code{D} and their subtypes are subtypes of the free
type \code{C<?>} corresponding to some class \code{C}, then \code{D}
\emph{is }a subclass of \code{C}.

As stated here, this statement is familiar to OO developers using
nominally-typed OO programming languages such as Java, C\#, C++, Scala
and Kotlin. It is a true statement in these languages due to the nominality
of subtyping in these languages. Subclassing (\emph{a.k.a.}, inheritance),
a relation characteristic of class-based OOP, is always specified
between classes in OO programs using class \emph{names}. Nominal subtyping
asserts a bidirectional correspondence between the subtyping relation
and the inherently nominal subclassing relation.

In the case of \emph{non}-generic OOP, the correspondence between
subtyping and subclassing is expressed, succinctly, by stating that
`inheritance \emph{is} subtyping'.. In the case of generic OOP,
the correspondence is succinctly expressed by stating that `inheritance
is \emph{the} source of subtyping'. The latter is a compact expression
of Equation~(\ref{eq:adj}).

Focusing on generic nominally-typed OOP, the `inheritance is the
source of subtyping' statement and Equation~(\ref{eq:adj}) state,
first (in the left-to-right direction), that subclassing does result
in (\emph{i.e.}, is \emph{a} source of) subtyping between reference
types, and, secondly (in the right-to-left direction), that subclassing
is the \emph{only }source of subtyping between reference types (\emph{i.e.},
that besides subclassing there are no other sources for subtyping).\medskip{}
} making erasure and free types in OO programming languages two adjoints
of an adjunction---the \emph{JEA} adjunction.\endnote{It should be noted that in Java and other similar OO languages a parameterized
type is always a subtype of the free type corresponding to the class
resulting from the erasure of the parameterized type, and a class
is always the same as that resulting from the erasure of the free
type corresponding to the class. In symbols, in OO generics we have
\[
t<:FT\left(E\left(t\right)\right)\textrm{ and }c=E\left(FT\left(c\right)\right)
\]
for all types $t$ and classes $c$.

For example, in Java we have \code{List<String>~<:~List<?>} and,
more generally, we have \code{List<T>~<:~List<?> = $FT$(List) = $FT$($E$(List<T>))}
for all types \code{T}. Also, for all classes \code{C} we have \code{C}
as the erasure of the corresponding free type \code{C<?>}, \emph{e.g.},
we have \code{LinkedList = $E$(LinkedList<?>) = $E$($FT$(LinkedList))}
and \code{List = $E$(List<?>) = $E$($FT$(List))}.

Following order-theoretic parlance~\cite[Ch. 7]{Davey2002}, a reference
type is called a \emph{closed type} if it does not get changed (\emph{i}.\emph{e.},
is fixed) by erasure followed by free type construction. Similarly,
a class is called a \emph{closed} \emph{class} if it is not affected
by free type construction followed by erasure. As such, relative to
\emph{JEA}, in the OO subtyping relation only the free types are closed
types, while in the subclassing relation all classes are closed classes.

In order theory parlance, the compositions $FT\circ E$ (from the
subtyping relation to itself) and $E\circ FT$ (from the subclassing
relation to itself) of the adjoints of \emph{JEA} are called \emph{closure
operators}, and closed elements are \emph{fixed} \emph{points} of
these closure operators~\cite[Ch. 7]{Davey2002}.

These observations regarding \emph{JEA} and closure operators hint
at the well-known correspondence between adjunctions and monads in
category theory. (See the discussion in the main text for another
use of monads in modeling generics.)\medskip{}
}

It should be noted that the notion of a free type in OOP is similar
to the notion of a `free monoid' corresponding to a set and of a
`free category' (a `quiver') corresponding to a graph. More details
on \emph{JEA} and on free types are available in~\cite{AbdelGawad2017b}
and~\cite{AbdelGawad2019b}.

\paragraph{Monads, Inductive/Coinductive Types, and Co-free Types}

Monads and comonads in category theory generalize closure and kernel
operators in order theory~\cite{Fong2018,Priestley2002}. As such,
the discussion of (co)induc\-tive types in the order-theoretic approach
to modeling Java generics~\cite{AbdelGawad2018f,AbdelGawad2019b}---a
discussion that can be expressed using closure and kernel operators~\cite{Davey2002}---can
also be generalized to involve categories (as ``generalized posets''),
using monads and comonads. Generalizing the discussion of in\-duction/co\-induc\-tion
to categories has the immediate benefit of enabling a discussion of
inductive and coinductive types while \emph{not} requiring the subtyping
relation to be a complete lattice, nor requiring the existence of
exact fixed points in the relation. (We further discuss this point
in~\cite{AbdelGawad2018f,AbdelGawad2019b}.)

Using order-theory, inductive types and coinductive types in OOP (which
are relevant to $F$-bounded generics\endnote{An example of $F$-bounded generics is class \code{Enum} in Java.
The declaration of class \code{Enum} looks as follows:

\code{~~~\textbf{\small{}class}{\small{} Enum<T }\textbf{\small{}extends}{\small{}
Enum<T>\textcompwordmark > \{...\}}}.

Class \code{Enum} is an instance of an $F$-bounded generic class
because its type parameter \code{T} is upper-bounded by a parameterized
type that uses a type variable---namely \code{T} itself---as a
type argument (Parameter \code{T} is thus not bounded by a constant
type, but rather by a function/functor over types---hence the name
`$F$-bounded'\footnote{By allowing the bound of an $F$-bounded type variable to depend on
\emph{any} type variable, not just on the bound type variable, we
are sticking to the original definition of $F$-bounded polymorphism
(which posits that `the bound type variable \emph{may} occur within
the bound'~\cite{CanningFbounded89}) rather than to the definition
that is seemingly used in more recent literature, \emph{e.g.}, in~\cite{Greenman2014}
(which posits that a type variable is $F$-bounded if and only if
the variable \emph{does} occur, recursively, in its own bound).}).

As such, the type parameter \code{T} of class \code{Enum} ranges
over all \emph{$F$-subtypes} of class \code{Enum} (the \emph{second}
occurrence of \code{Enum} in the declaration), \emph{i.e.}, over
all \code{Enum}-subtypes. (The $F$-subtypes of class \code{Enum}
are also called the \emph{coinductive} or \emph{post-fixed} types
of \code{Enum}, or \code{Enum}\emph{-coinductive} or \code{Enum}\emph{-postfixed}
types).

In summary, \emph{valid }type arguments to class \code{Enum} are
only \code{Enum}-subtypes (\emph{i.e.}, of all the admittable parameterized
types constructed using class \code{Enum} the \emph{valid} ones are
only those instantiations of \code{Enum} with \code{Enum}-subtypes.
For a discussion of admittable type arguments versus valid ones, of
admittable parameterized types---such as \code{Enum<Object>}---versus
valid ones, and of admittable subtyping relations versus valid ones,
see~\cite{AbdelGawad2018e}.)\medskip{}
}%
) are modeled using pre-fixed points and post-fixed points. When using
monads from category theory, the discussion of inductive and coinductive
types can be expressed, rather, using the category-theoretic notions
of $F$\nobreakdash-algebras and $F$\nobreakdash-coalgebras%
. In particular, if $F$ is a generic class, then all the parameterized
types \code{Ty} such that \code{Ty} is a subtype of \code{F<Ty>}
are called the $F$\emph{-subtypes} of class $F$, while, dually,
all the parameterized types \code{Ty} such that \code{F<Ty>} is
a subtype of \code{Ty} are called the $F$\emph{-supertypes} of $F$.
Using monads enables easily seeing that $F$-subtypes of a generic
class $F$ in an OO program directly correspond to coalgebras of $F$,
while its $F$-supertypes correspond to algebras of $F$.

Using monads further allows easily seeing that free types, as the
greatest $F$-subtypes, are \emph{final} coalgebras in the Java subtyping
category%
, and that, on the other hand, initial algebras rarely exist in the
Java subtyping relation since, unlike for free types, Java does \emph{not}
define a general notion of types that correspond to least $F$-supertypes.
The discussion of final coalgebras and initial algebras motivates
us to suggest adding to Java and similar OOP languages the notion
of \emph{co-free types} as the least $F$-supertypes (\emph{i.e.},
as initial algebras) in the OO subtyping relation. Co-free types,
as indicated by their name, function as duals of free types. We envision
the main use of co-free types to be as lower bounds of type variables
and of interval type arguments.\endnote{In particular, for each generic class \code{C}, we suggest the notation
\code{C<!>} for the corresponding co-free type. In Java the cofree
type \code{C<!>} has as instances \emph{only} the trivial object
\code{null}. The cofree type \code{C<!>} has as supertypes all parameterized
types that are instantiations of class \code{C} (and their supertypes),
and has as subtypes only the cofree types corresponding to all subclasses
of \code{C}. 

We predict the main use of cofree types to be as lower bounds of type
variables (in doubly $F$-bounded generics~\cite{AbdelGawad2018e})
and of interval type arguments (in interval types~\cite{AbdelGawad2018c}).
Currently in Java, rather confusingly, when a free type such as \code{C<?>}
is used as the lower bound of a wildcard type argument\emph{ }(\emph{e.g.},
as in \code{?~super~C<?>}) the free type has a meaning---one similar
to that of co-free types---that is intuitively \emph{different }than
the meaning of the free type when it is used as the upper bound of
a wildcard type argument (\emph{e.g.}, as in \code{?~extends~C<?>})---which
hints at the need for co-free types.\medskip{}
} More details on the uses of lower bounds in OO generics can be found
in~\cite{AbdelGawad2018c,AbdelGawad2018e}.

\paragraph{\emph{JSO}: An Operad for Constructing the Generic Subtyping Relation}

Operads are a tool in category theory that can be used to model self-similar
phenomena~\cite{spivak2014category}. As such, operads can be used
to construct the subtyping relation in generic OOP. Noting the self-similarity
of the Java subtyping relation allowed us, in our order-theoretic
approach to modeling generics, to express the construction process
of the subtyping relation \emph{iteratively}~\cite{AbdelGawad2017a,AbdelGawad2018b}.
In~\cite{AbdelGawad2017a}, in particular, we presented an outline
for defining $JSO$ (the Java Subtyping Operad) as an operad that
models this iterative construction process. We expect the full development
of \emph{JSO }to be a significant step in the construction of a comprehensive
category-theoretic model of OO generics.

\paragraph{Discussion}

Based on the substantial progress we made in developing an order-theoretic
and category-theoretic approach to modeling generics, the concepts
and tools of order theory and category theory seem to us as being
perfectly suited for modeling generics in nominally-typed OOP languages.

As such we believe order theory and category theory hold the keys
to overcoming the challenges met when modeling generics, and thus
also to developing a simple, yet accurate model of generics in mainstream
object-oriented programming languages such as Java, C\#, C++, Scala
and Kotlin.\endnote{\emph{Collaboration Plea: Given the limited human and financial resources
available to him, the author would like to invite interested and qualified
parties }(\emph{such as individual mathematicians and PL researchers,
PL and theoretical computer science research groups, software-development
companies and corporations, research funding agencies, and other interested
parties})\emph{ to join him in speeding up the development of the
order-/cat\-egory-theo\-retic approach to modeling OO generics.}}

\bibliographystyle{plain}

\appendix
\theendnotes
\end{document}